# Multiple Softening *Q*-vectors Driving a Cascade of CDW Phases in $VSe_2$

*Zheng-Hong Li*[1], *Yung-Ting Lee*[2]*, *Yu-Chan Tai*[1], *Cheng-Tien Chiang*[3,4,5,6],
*Chien-Cheng Kuo*[5], *Meng-Kai Lin*[6], *Chun-Liang Lin*[1], *Hung-Chung Hsueh*[8]*,
*Ming-Chiang Chung*[9]*, *Po-Tuan Chen*[10]*, *Chi-Cheng Lee*[8]*

[1]*Department of Electrophysics, National Yang Ming Chiao Tung University, Hsinchu 300, Taiwan*

[2]*Department of Applied Science, National Taitung University, Taitung 950309, Taiwan*

[3]*Institute of Atomic and Molecular Sciences, Academia Sinica, Taipei 106, Taiwan*

[4]*Department of Physics, National Taiwan University, Taipei, 106, Taiwan*

[5]*Department of Physics, National Sun Yet-Sen University, Kaohsiung 804, Taiwan, Taiwan*

[6]*Department of Physics, National Central University, Taoyuan 32001, Taiwan*

[7]*Physics Division, National Center for Theoretical Sciences, Taipei, 106, Taiwan*

[8]*Department of Physics, Tamkang University, New Taipei City 251, Taiwan*

[9]*Department of Physics, National Chung Hsing University, Taichung 402, Taiwan*

[10]*Department of Vehicle Engineering, National Taipei University of Technology, Taipei 106, Taiwan*

## Corresponding Authors

E-mail: cclee@mail.tku.edu.tw (CCLee), chenpotuan@ntut.edu.tw (PTChen), ytl@nttu.edu.tw (YTLee), mingchiangha@phys.nchu.edu.tw (MCChung), hchsueh@gms.tku.edu.tw (HCHsueh).

**Abstract**

Charge density wave (CDW) formation in two-dimensional materials is governed by complex competing lattice instabilities that remain incompletely understood. Here, we investigate the structural evolution of monolayer 1T-$VSe_2$ using first-principles electronic and phonon calculations. The pristine phase exhibits several imaginary-frequency phonon modes associated with dominant instability wave vectors $\boldsymbol{Q}_{CDW}$, which generate the first-generation CDW phases. Subsequent phonon analyses reveal that several of these intermediate structures remain dynamically unstable and undergo further symmetry-lowering distortions into larger superstructures. Through iterative phonon-driven relaxations, we identify multiple transformation pathways that converge toward the same low-energy $2\sqrt{3} \times 4$ CDW configuration. Although these pathways originate from distinct intermediate CDW states, they ultimately reach nearly degenerate energetically stable phases, demonstrating that different phonon-driven routes can lead to the same ground-state configuration. The results establish a unified phonon-driven cascade mechanism for hierarchical CDW formation in monolayer 1T-$VSe_2$ and provide a systematic framework for understanding competing ordered phases in low-dimensional quantum materials.

**Keyword**

Charge Density Wave; Monolayer 1T-$VSe_2$; Phonon Instability; Density Functional Theory; Lattice Reconstruction.

## 1. Introduction

Charge density wave (CDW) is a collective quantum state characterized by periodic modulation of electron density accompanied by lattice distortion **[1,2]**. Because of its strong electron–phonon coupling and emergent collective behavior, the CDW state provides a promising platform for manipulating electronic order parameters in next-generation quantum technologies, including ultrafast switches, quantum memories, and low-dimensional quantum electronic devices **[3,4]**.

In one-dimensional systems, CDWs are commonly described by the Peierls model, in which Fermi-surface nesting induces lattice distortion and opens an energy gap at the Fermi level ($E_F$) **[1,2,5]**. However, the applicability of this mechanism to two-dimensional (2D) materials remains under debate [6]. Recently, Lee *et al*. demonstrated through first-principles calculations that multiple CDW-related structures in 2D 2H-$NbSe_2$ can be systematically connected through a hierarchy of lattice instabilities **[7]**. In that system, the filled phase represents the lowest-energy configuration, exhibiting energy gaps at the CDW Brillouin-zone boundary and a strongly suppressed density of states at $E_F$. Importantly, the stable CDW phase was shown to emerge through sequential activation of nesting vectors, involving higher-energy intermediate phases such as the stripe phase. These results suggest that the formation of the ground-state CDW is not a single-step transition, but rather a multistep evolution through causally connected metastable phases.

From a lattice-dynamical perspective, phonon softening governs the evolution from the normal state to the stable CDW phase. Imaginary phonon frequencies indicate unstable lattice distortions that initially generate intermediate metastable structures, such as stripe or unidirectional CDW orders **[8]**. These distortions reconstruct the electronic structure and Fermi surface, subsequently activating additional nesting vectors and secondary phonon instabilities. Further condensation of soft phonon modes lowers both symmetry and total energy, eventually producing the dynamically stable ground-state CDW phase with reduced density of states at $E_F$. This physical picture highlights that CDW formation is controlled by coupled electronic and lattice instabilities through a cascade of interconnected metastable states. Although such a mechanism is likely applicable to many 2D transition-metal dichalcogenides (TMDs), computationally capturing the entire evolution remains challenging because density functional theory (DFT) calculations are computationally expensive, while molecular dynamics simulations often lack sufficient accuracy for CDW energetics **[9,10]**. Consequently, only limited studies have fully described the complete transition pathway.

Among 2D TMDs, monolayer $VSe_2$ has attracted considerable attention because of its rich and highly tunable CDW phenomena **[11-13]**. Previous studies have reported multiple competing CDW phases, together with several intermediate or metastable distortions associated with metal–insulator transitions and anharmonic lattice effects **[13,14]**. Experimental investigations further revealed that the CDW order in monolayer $VSe_2$ is strongly affected by strain, substrate coupling, dimensional confinement, and electron–phonon interactions. Despite the large number of reported CDW structures, most previous works treated these phases as isolated states, while the causal relationships and transformation pathways among them have rarely been explored. However, clarifying the evolutionary connections between the normal phase, intermediate metastable states, and the final stable phase is essential for understanding competing orders and exploiting CDWs as functional elements in quantum devices.

In this work, we investigate the phonon-mediated cascade mechanism to the intricate and controversial CDW phases of monolayer $VSe_2$. Through successive phonon softening processes, including previously unidentified intermediate structures, we demonstrate that many experimentally or theoretically reported CDW phases remain dynamically unstable because of substantial residual imaginary phonon modes, indicating that they do not correspond to the true ground state. By following the complete soft-mode-driven evolution pathway, we further identify a new CDW phase that represents the genuine lowest-energy configuration of the system.

## 2. Method

### *2.1. DFT computational details*

All first-principles calculations for 1T-$VSe_2$ were performed using the OpenMX package within density functional theory (DFT) **[15,16]**. The in-plane lattice constants were fully relaxed, yielding $a = b = 3.29$ Å with $\gamma = 120^{\circ}$. The exchange–correlation functional was treated within the generalized gradient approximation (GGA) using the Perdew–Burke–Ernzerhof (PBE) parametrization **[17]**. Norm-conserving pseudo-potentials and optimized pseudo-atomic basis functions were employed, with V6.0-*s3p2d1* and Se7.0-*s3p2d2*, where the numbers 1, 2, and 3 indicate the number of orbitals and 6.0 and 7.0 implies to cutoff radius of the pseudo-atomic orbital **[18,19]**.

A real-space grid energy cutoff of 1000 Ry was used to ensure numerical convergence. Notably, this cutoff is higher than typically employed, as our tests indicated that such a value is required to achieve sufficiently accurate structural energies and phonon properties. Structural relaxations were performed until residual forces were below $1 \times 10^{-4}$ Hartree/Bohr and total energy convergence criteria reached $1 \times 10^{-9}$ Hartree. The primitive-cell Brillouin zone was sampled using a $42 \times 42 \times 1$ Monkhorst–Pack grid **[20]**, and comparable ***k***-point meshes were employed for the supercell calculations. Structural visualizations were produced using VESTA **[21]** and OpenMX viewer **[22].**

Phonon dispersions were calculated using the finite-displacement method (FDM) by adopting the supercells commensurate with the wavelengths **[23,24]**. Lattice instabilities were identified from imaginary-frequency phonon modes **[25]**, which appear on the phonon instability map (PIM) with frequencies below 10 cm$^{-1}$.

*2.2. Phonon-driven structural evolution*

The CDW ordering vector $\boldsymbol{Q}_{CDW}$ was identified from the wavevector of the phonon softening **[26,27]**. Symmetry-breaking lattice distortions were constructed from the unstable mode eigen-displacements and fully relaxed in the corresponding supercells. An iterative phonon-driven search was carried out starting from the normal phase. Unstable phonon modes were used to generate distorted structures, which were subsequently relaxed and examined by FDM to identify any residual instabilities. As illustrated in **Fig. 1**, the normal phase is located at the top of the energy landscape, while the intermediate states correspond to saddle points or metastable states, depending on whether phonon soft modes are still present. Ultimately, the search leads to the most stable state.

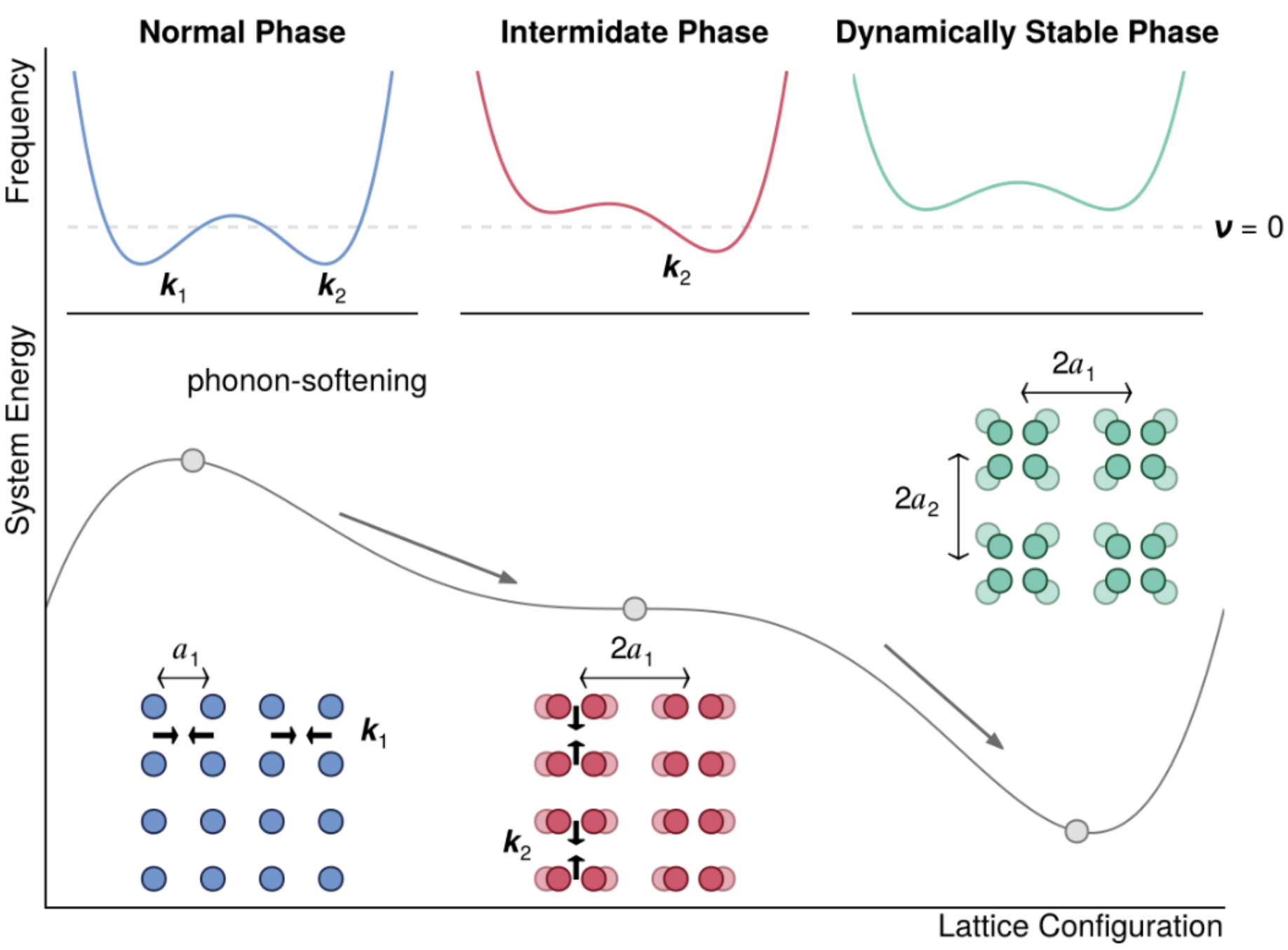

**Figure 1.** Schematic illustration of the phonon-driven cascade. The top panels show the phonon dispersions of the normal phase, intermediate CDW phase, and dynamically stable CDW phase, while the bottom panel illustrates the lattice distortions in the CDW phases. The system evolves via phonon softening from the normal phase to the intermediate CDW phase, and finally to the dynamically stable CDW phase, which is dynamically stable.

This hierarchical procedure captures successive CDW phases arising from the sequential condensation of soft modes. The real-space modulation period is determined by the reciprocal space ordering vector $\boldsymbol{Q}_{CDW}$, and the supercells size are chosen to be commensurate with the wavelength λ, defined by

$$\lambda = 2\pi / |\boldsymbol{Q}_{CDW}|. \tag{1}$$

Each unstable mode produces a structure with reduced symmetry, which may contain additional soft phonon modes. By iteratively relaxing these instabilities phonon mode as shown on **Fig.1**, the atomic structure evolves through a cascade of symmetry-breaking transitions, forming a hierarchy of CDW phases. Each step is determined by the dynamical matrix eigenvectors which defined by

$$D_{i\alpha,j\beta}(q) = \sum_j \frac{\Phi_{i\alpha,j\beta}}{\sqrt{M_i M_j}} e^{iq\cdot(R_j - R_i)}, \tag{2}$$

providing a self-consistent mapping of the instability landscape. where i and j denote atomic indices, with $i \neq j$, while α and β indicate the Cartesian components (x, y, z). Here, M represents the atomic mass, R the atomic position vector, and Φ the force-constant matrix element **[25]**. The resulting dynamical matrix enables a self-consistent characterization of the instability landscape. The iteration terminates when a dynamically stable structure, free of imaginary frequencies, is reached, thereby identifying all accessible CDWs configurations within this lattice-dynamical framework.

## 3. Results and discussion

*3.1. Phonon instabilities of the pristine monolayer $VSe_2$*

Several CDW phases have previously been reported for monolayer $VSe_2$, including the $4 \times 1$, $\sqrt{3} \times 4$, and $\sqrt{3} \times \sqrt{7}$ superstructures **[11,12,26-29]**. **Fig. 2(a)** shows the phonon dispersion of the pristine 1×1 monolayer 1T-$VSe_2$ along the high-symmetry path Γ–M–K–Γ. Pronounced imaginary-frequency phonon modes are observed, indicating strong dynamical instability of the normal phase. These unstable modes occur at three characteristic wave vectors, labeled as $\boldsymbol{Q}_A$ = (1/5, 1/5, 0), $\boldsymbol{Q}_B$ = (1/8, 1/8, 0), and $\boldsymbol{Q}_C$ = (1/4, 0, 0). To visualize the distribution of these unstable phonon modes throughout the Brillouin zone, the imaginary phonon frequencies are further represented by a phonon instability map, as shown in **Fig. 2(b)**. This representation provides a comprehensive view of the unstable regions in the entire reciprocal space and enables systematic identification of all lattice instabilities.

Mapping the wave vectors $\boldsymbol{Q}_A$, $\boldsymbol{Q}_B$, and $\boldsymbol{Q}_C$ from reciprocal space to real-space lattice distortions reveals the corresponding structural modulations. The real-space phonon eigen-displacements are illustrated in **Fig. 2(c)**, where the atomic displacement patterns associated with the unstable modes generate three primary CDW superstructures: $\sqrt{3} \times \sqrt{7}$, $\sqrt{3} \times 4$, and $4 \times 1$ Phase. These structures therefore represent the first-generation CDW phases emerging directly from the phonon instability of the normal phase.

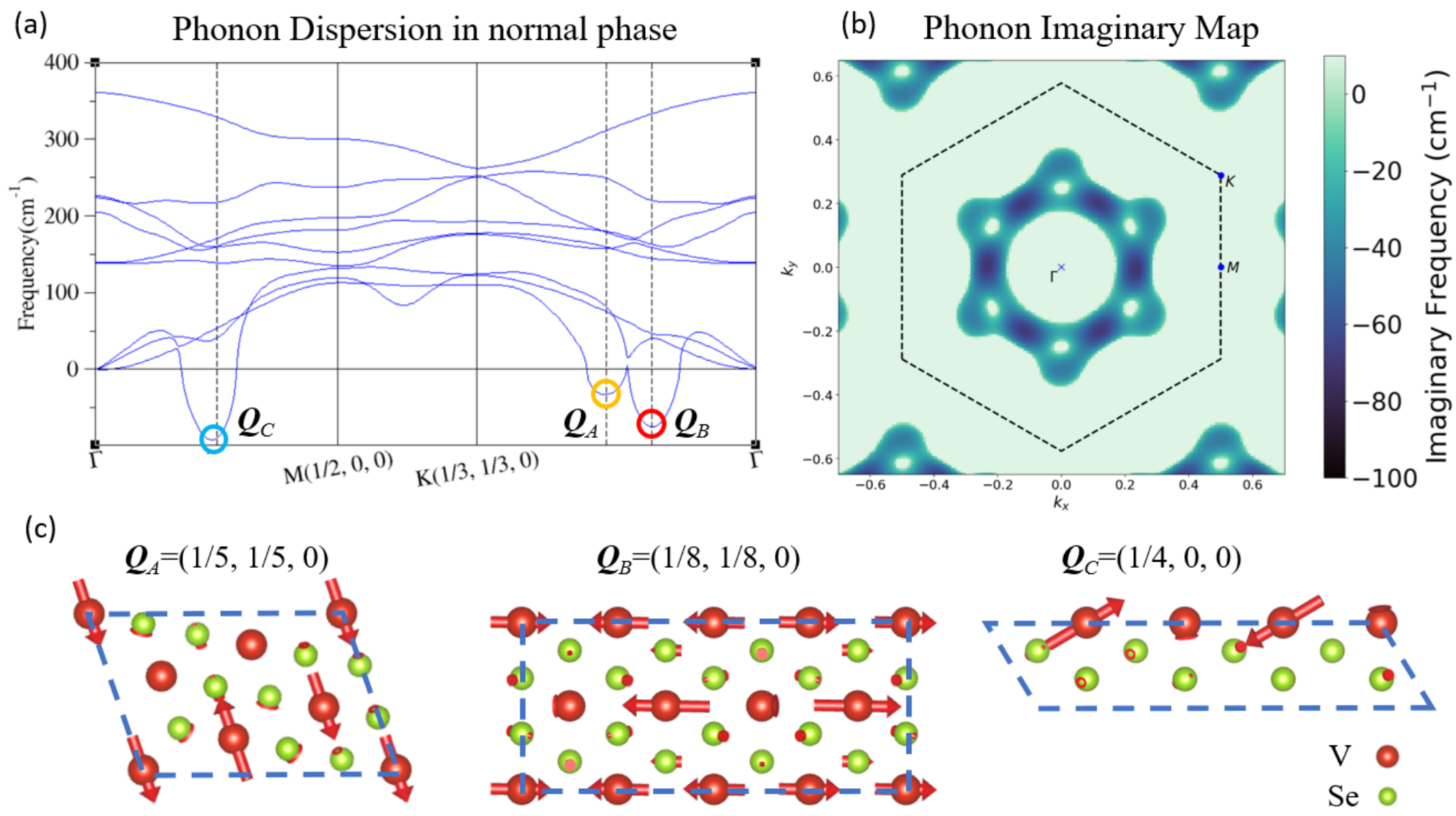

**Figure 2.** (a) Phonon spectrum in normal phase 1T-$VSe_2$, (b) map of imaginary phonon frequencies, and (c) eigen displacement vectors of the unstable modes in the monolayer $1 \times 1$ pristine 1T-$VSe_2$. Imaginary-frequency modes at wave vectors $\boldsymbol{Q}_A$ = (1/5, 1/5, 0), $\boldsymbol{Q}_B$ = (1/8, 1/8, 0), and $\boldsymbol{Q}_C$ = (1/4, 0, 0) reveal the primary lattice instabilities that drive a phonon-mediated cascade of CDW phases.

### *3.2. Secondary phonon instabilities of the CDW phases*

To evaluate the dynamical stability of these first-generation CDW structures, phonon dispersions were calculated for the relaxed $\sqrt{3} \times \sqrt{7}$, $\sqrt{3} \times 4$, and $4 \times 1$ phases. Their phonon instability maps and the corresponding spatial displacements of soft phonon modes for the lattice two phases at $\boldsymbol{Q}_{B1}$, $\boldsymbol{Q}_{B2}$, $\boldsymbol{Q}_{C1}$, and $\boldsymbol{Q}_{C2}$ are shown in **Fig. 3 (a)**.

Among these structures, the $\sqrt{3} \times \sqrt{7}$ phase exhibits no remaining imaginary phonon modes, indicating that this structure is dynamically stable. In contrast, additional lattice instabilities are still observed in the other two phases. The $\sqrt{3} \times 4$ structure exhibits two unstable phonon modes at $\boldsymbol{Q}_{B1}$ = (1/2, 0, 0) and $\boldsymbol{Q}_{B2}$ = (0, 1/2, 0) while the $4 \times 1$ structure shows two unstable modes located at $\boldsymbol{Q}_{C1}$ = (-1/2, 1/4, 0) and $\boldsymbol{Q}_{C2}$ = (0, 1/4, 0). These imaginary-frequency phonon maps indicate that both of the $\sqrt{3} \times 4$ and $4 \times 1$ phases remain dynamically unstable. The associated phonon eigen-displacements reveal the directions of further structural distortions that drive the next stage of the phase cascade.

Guided by these eigen-displacements, further structural relaxations were then performed. The resulting lattice distortions give rise to four secondary CDW structures: $2\sqrt{3} \times 4$, $\sqrt{3} \times 8$, $4 \times 2\sqrt{3}$, and $4 \times 4$ phases, as illustrated in **Fig. 3(b)**. Specifically, following the distortion indicated with $\boldsymbol{Q}_{C1}$ drives the $4 \times 1$ phase evolves into the $4 \times 4$ configuration. In contract, the $4 \times 1$ phase associated with $\boldsymbol{Q}_{C2}$ progresses toward the $4 \times 2\sqrt{3}$ configuration, while the $\sqrt{3} \times 4$ phase, governed by $\boldsymbol{Q}_{B1}$, independently relaxes toward the closely related $2\sqrt{3} \times 4$ and $\sqrt{3} \times 8$ structure. Finally, the $\sqrt{3} \times 8$ phase can relax into the closely related $2\sqrt{3} \times 8$ phase, as showed on **Fig.4**. The nearly identical energies of these configurations, as listed in Table 1, demonstrate that distinct metastable CDW pathways converge to the same low-energy CDW state within the phonon-driven cascade.

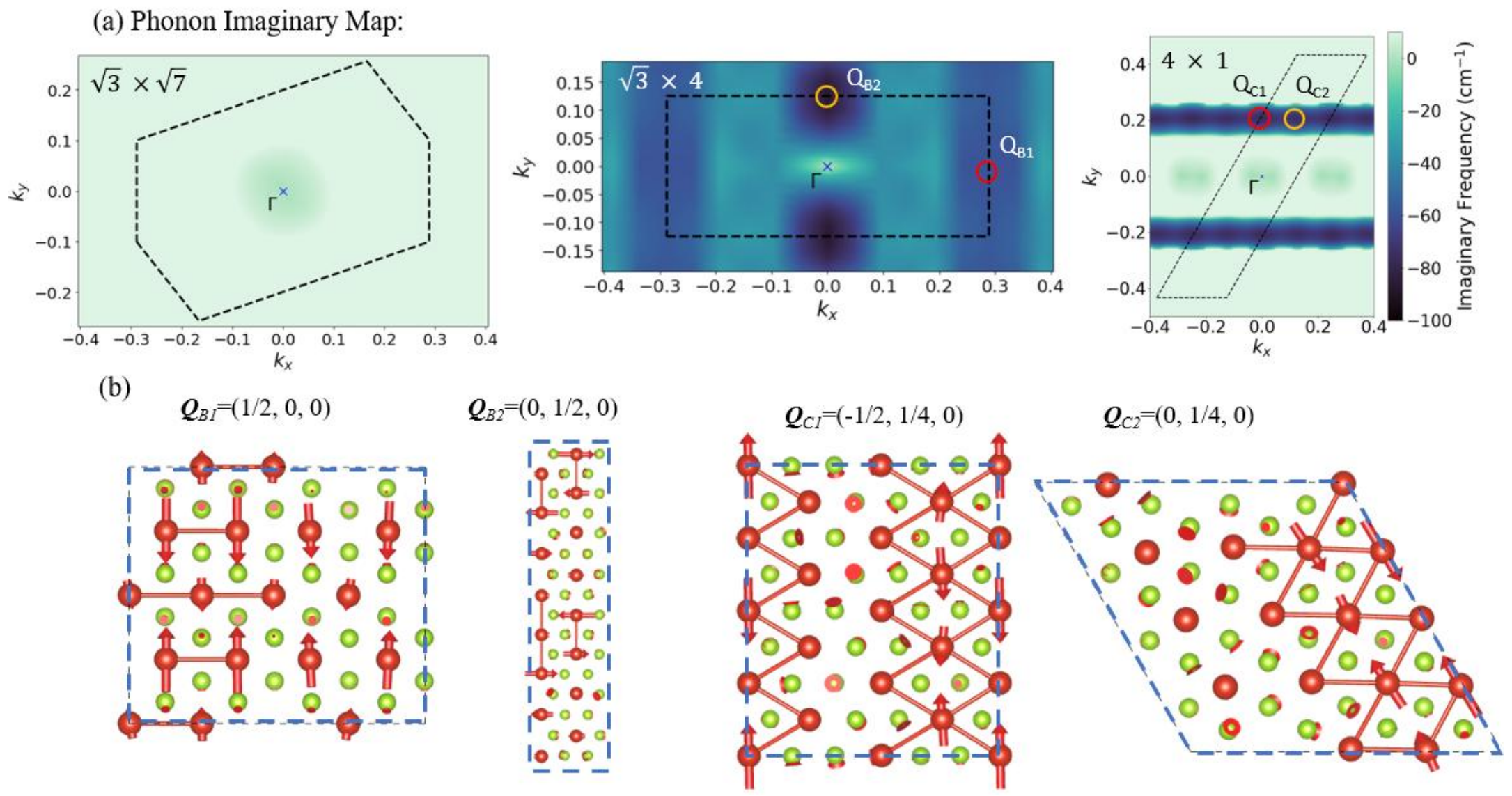

**Figure 3.** Soft phonon modes and corresponding eigen-displacements of the first-generation of the CDW phases. (a) Imaginary-frequency map of phonon spectra of the relaxed $2\sqrt{3}\times 4$, $\sqrt{3}\times 8$, $4\times 4$ and $4\times 2\sqrt{3}$ supercell structures. Imaginary-frequency phonon modes (deep blue regions) reveal remaining lattice instabilities. (b) The associated real-space soft phonon eigen-displacements indicate the structural distortions responsible for the next stage of the phase cascade, including the formation of $2\sqrt{3}\times 4$, $\sqrt{3}\times 8$, $4\times 4$ and $4\times 2\sqrt{3}$ superstructures.

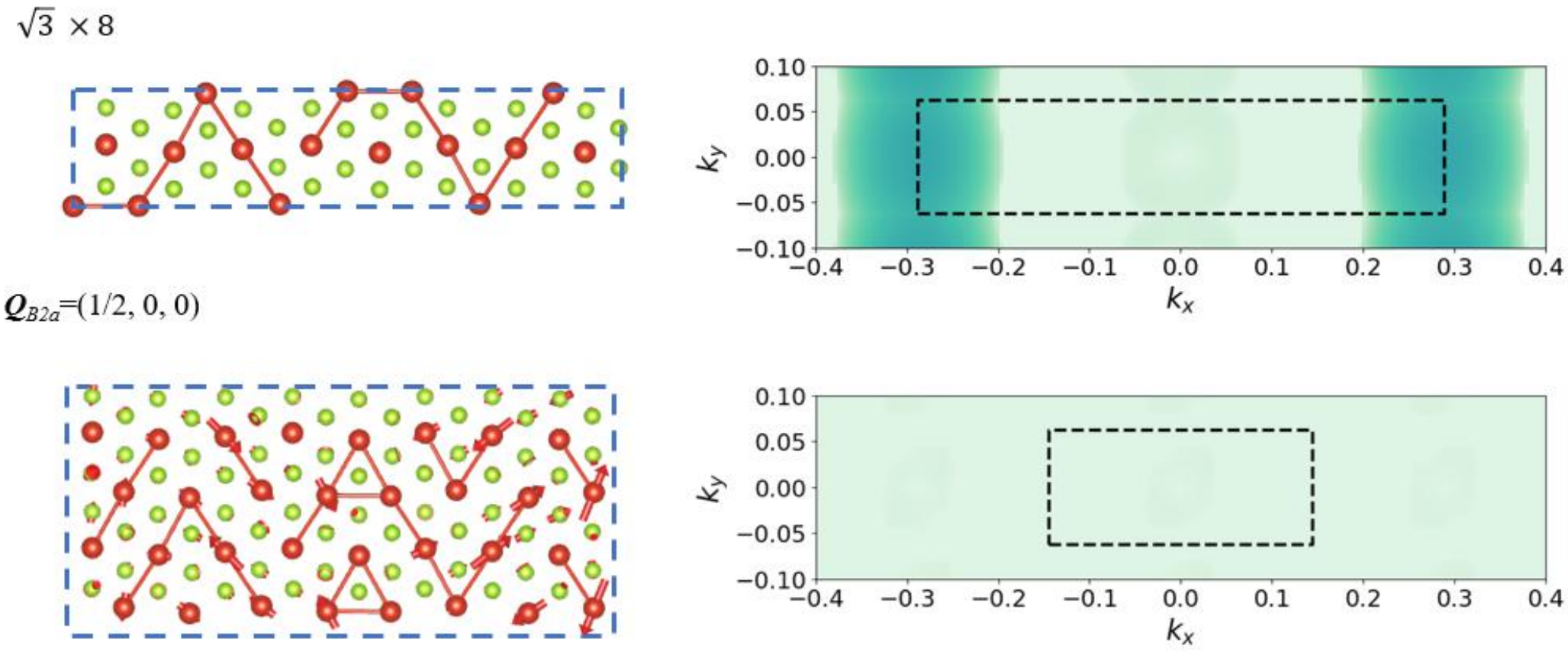

**Figure 4.** Soft phonon modes and corresponding eigen-displacements of the first-generation the $\sqrt{3}\times 8$ CDW phases. Imaginary-frequency map of phonon spectra of the relaxed $\sqrt{3}\times 8$, and $2\sqrt{3}\times 8$ supercell structures. Imaginary-frequency phonon modes (deep blue regions) reveal remaining lattice instabilities.

**Table 1.** Structural parameters and relative energies of the CDW phases of monolayer 1T-$VSe_2$.

| Phase | Wavevector | ΔE (meV/per atom) | Theo. | Exp. |
|---|---|---|---|---|
| $1 \times 1$ | - | 0.00 | **[26,27]** | **[14]** |
| $\sqrt{3} \times \sqrt{7}$ | $Q_A = (1/5, 1/5, 0)$ | -1.49 | **[13,26,27]** | **[27,14]** |
| $\sqrt{3} \times 4$ | $Q_B = (1/8, 1/8, 0)$ | -0.42 | **[13]** | **[30]** [a] |
| $4 \times 1$ | $Q_C = (1/4, 0, 0)$ | -1.62 | **[13]** | - |
| $2\sqrt{3} \times 4$ | $Q_{B1} = (0, 1/2, 0)$ | -2.03 | This work | - |
| $\sqrt{3} \times 8$ | $Q_{B2} = (1/2, 0, 0)$ | -1.49 | This work | - |
| $4 \times 2\sqrt{3}$ | $Q_{C1} = (1/2, 1/4, 0)$ | -2.02 | This work | - |
| $4 \times 4$ | $Q_{C2} = (0, 1/4, 0)$ | -1.92 | **[13,26]** | **[31,32]** [b] |
| $2\sqrt{3} \times 8$ | $Q_{B2a} = (0, 1/2, 0)$ | -1.88 | This work | **-** |

[a] The experimentally observed $\sqrt{3} \times 4$ structure corresponds to monolayer $VSe_2$ grown on $Al_2O_3$, where strain was included in the accompanying DFT calculations.

[b] The $4 \times 4$ phase has been identified in STM measurements of bilayer and bulk $VSe_2$.

### *3.3. Energetic Convergence of Phonon-Driven CDW*

The dynamical stability of these secondary CDW structures, $4 \times 4$ and $4 \times 2\sqrt{3}$ supercells, as shown in **Fig. 5(a) and (c)**, was further examined by calculating their phonon spectra. The resulting phonon imaginary-frequency maps are shown in **Fig. 5(b) and (d)**.

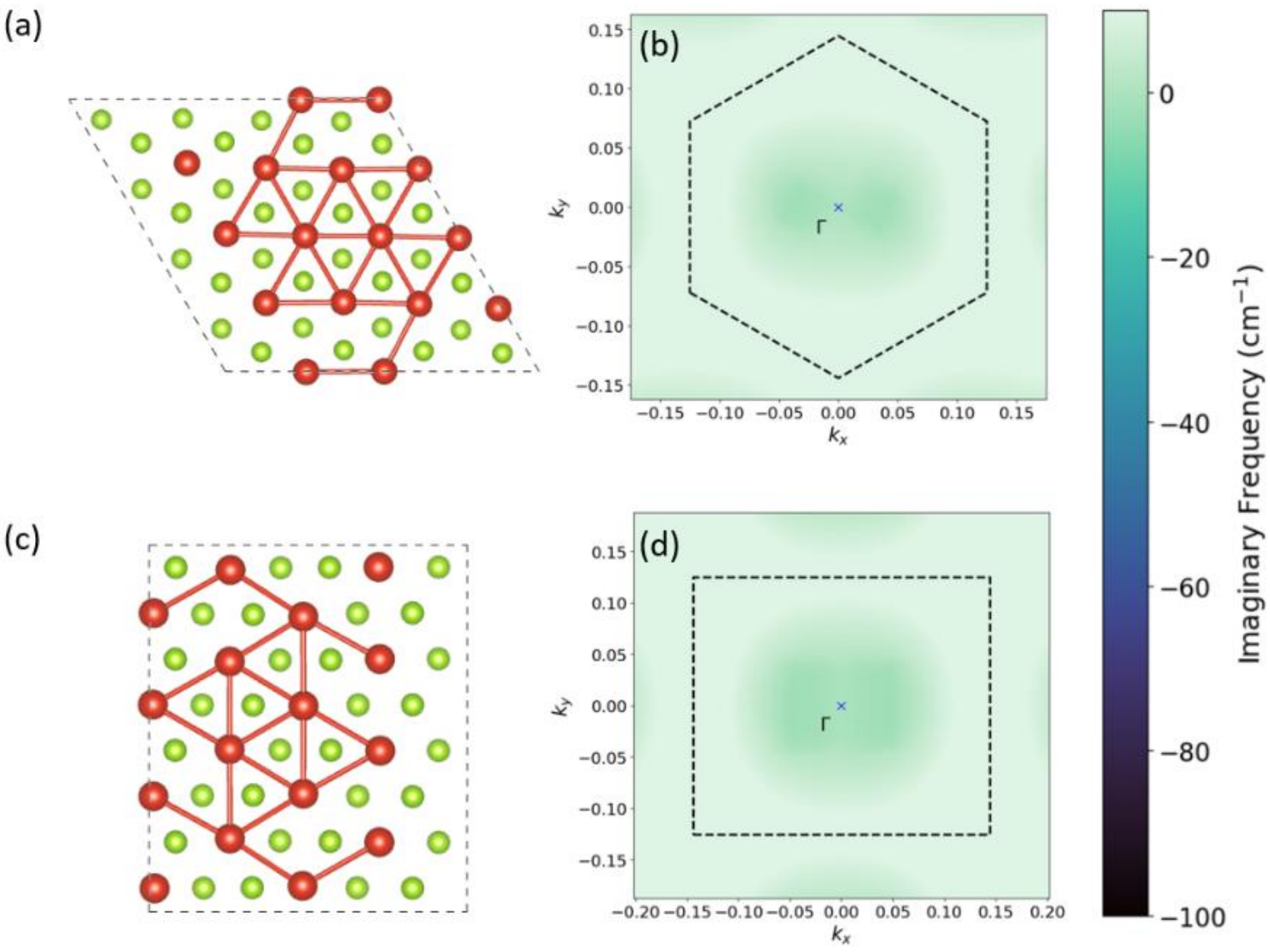

**Figure 5.** (a) and (b) are the top view of optimized $4 \times 4$ supercell structure of the CDW phase and corresponding imaginary frequency of phonon spectra in the Brillouin zone. (c) and (d) are the top view of optimized $4 \times 2\sqrt{3}$ supercell structure of the CDW phase and corresponding imaginary frequency of phonon spectra in the Brillouin zone. No imaginary phonon modes are observed in the entire Brillouin zone, demonstrating the dynamical stability of this configuration.

The phonon calculations of $4 \times 4$ and $4 \times 2\sqrt{3}$ structures show that the newly generated structures no longer exhibit imaginary-frequency phonon modes, indicating that these phases correspond to dynamically stable configurations. This result demonstrates that the system evolves through successive phonon-driven distortions until stable CDW phases are reached. To further evaluate the relative stability of these structures, the total energies and lattice parameters of all investigated CDW phases are summarized in **Table 1**.

Starting from the pristine $1 \times 1$ structure, the three primary phonon instabilities produce the $4 \times 1$, $\sqrt{3} \times \sqrt{7}$, and $\sqrt{3} \times 4$ CDW structures with energy reductions of $-1.62$, $-1.49$, and $-0.42$ meV per atom, respectively. Among these first-generation phases, the $4 \times 1$ structure exhibits the largest energy lowering, suggesting that Among these first-generation phases, the $4 \times 1$ structure exhibits the greatest energy gain, indicating that the lattice modulation associated with the $Q_C$ vector is more stable than those induced by $Q_A$ and $Q_B$. Further phonon-driven transformations from these intermediate structures further stabilize the energy. In particular, the $\sqrt{3} \times 4$ phase evolves into the $2\sqrt{3} \times 4$ and $\sqrt{3} \times 8$ structure with a total energy reduction of −2.03 and -1.49 meV per atom, while the $4 \times 1$ phase transforms into the $4 \times 4$ and $4 \times 2\sqrt{3}$ structures with energies of −1.92 and −2.02 meV per atom, respectively. Finally, $\sqrt{3} \times 8$ relax into $2\sqrt{3} \times 8$ structures with energies of -1.88 meV per atom.

These results show that the secondary CDW phases are consistently lower in energy than their parent structures, verifying that the first-generation CDW phases correspond to metastable intermediates in the phonon-driven structural transformation pathway. The phonon-driven cascade reveals that different structural pathways ultimately converge toward the closely related low-energy configurations. In particular, the $4 \times 1$ phase evolves toward the $4 \times 2\sqrt{3}$ structure through phonon-driven distortions, while the $\sqrt{3} \times 4$ phase independently transforms into the $2\sqrt{3} \times 4$ $\sqrt{3} \times 8$ configuration, which can further relax into the $2\sqrt{3} \times 8$ phase.

Despite originating from different intermediate structures, these final configurations exhibit nearly identical total energies (−2.02 and −2.03 meV per atom below the normal phase). This convergence indicates that multiple competing lattice instabilities guide the system toward a common low-energy region of the CDW energy landscape. The overall structural evolution can therefore be described as a phonon-driven cascade, in which each intermediate CDW phase retains residual lattice instabilities that condense and drive further reconstruction toward the true ground state, as illustrated schematically in **Fig. 6**.

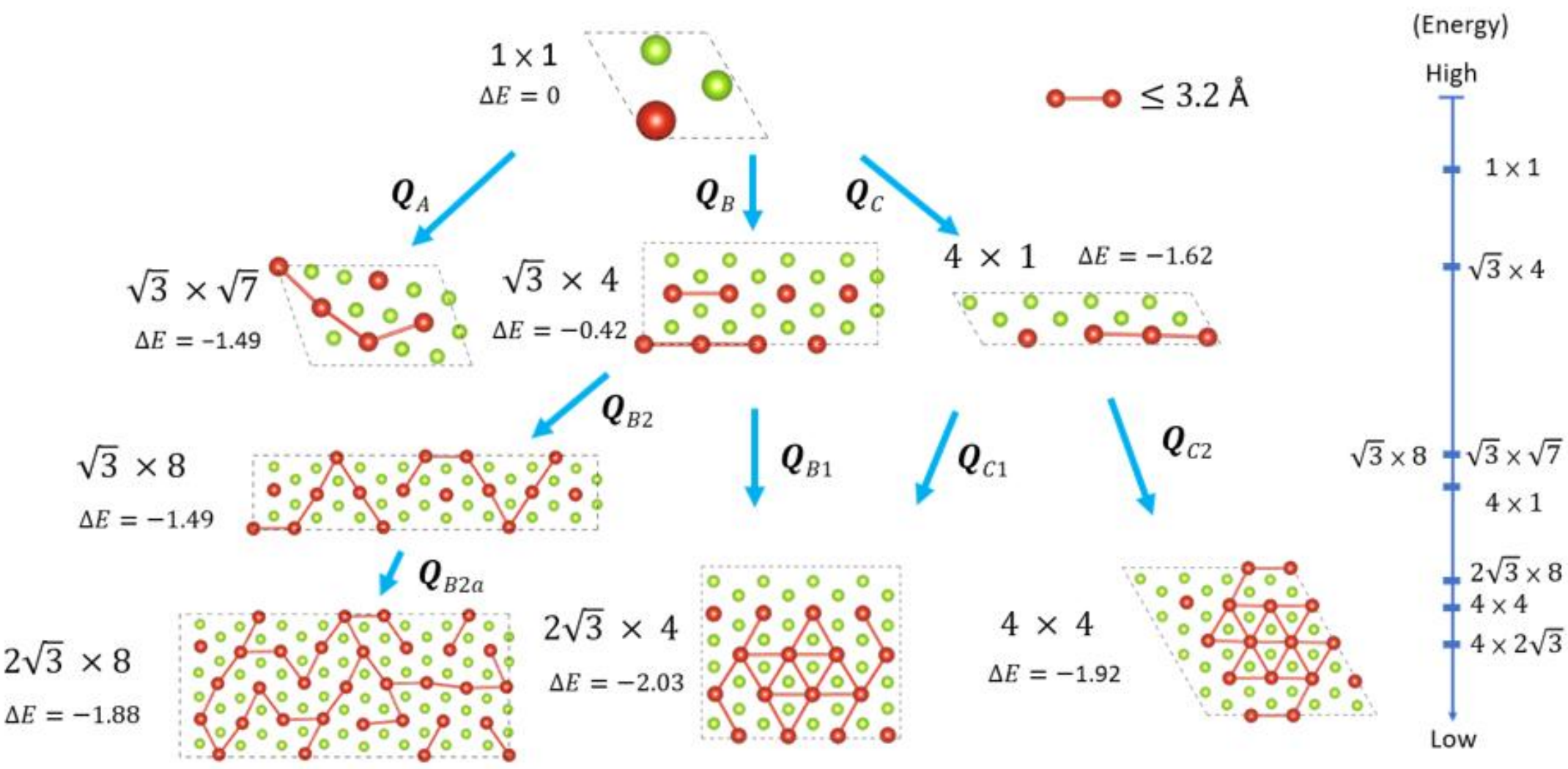

**Figure 6.** Phonon-driven CDW phase cascade in monolayer 1T-$VSe_2$. Schematic illustration of the structural evolution driven by phonon instabilities starting from the normal $1 \times 1$ phase. The initial phonon softening leads to several first-generation CDW configurations, including the $\sqrt{3} \times \sqrt{7}$, $\sqrt{3} \times 4$, and $4 \times 1$ phases. Subsequent lattice instabilities drive further structural transformations that ultimately converge toward the $4 \times 2\sqrt{3}$ phase, which is identified as the dynamically stable configuration within the explored phase space. Note that when V-V bond lengths are less than or equal to 3.2 angstroms, V-V bonds are represented by red stickers.

## 4. Summary

In summary, we have established a phonon-driven cascade framework to investigate the charge density wave (CDW) formation in monolayer 1T-$VSe_2$. Starting from the pristine 1×1 phase, the unstable phonon modes at $\boldsymbol{Q}_A$, $\boldsymbol{Q}_B$, and $\boldsymbol{Q}_C$ generate the first-generation CDW structures, including the $\sqrt{3} \times \sqrt{7}$, $\sqrt{3} \times 4$, and $4 \times 1$ phases.

By performing iterative phonon analysis on these relaxed CDW phases, we show that the $\sqrt{3} \times 4$, and $4 \times 1$ structures remain dynamically unstable and further transform into lower-energy configurations, whereas the $\sqrt{3} \times \sqrt{7}$ phase is already dynamically stable. The subsequent phonon-driven relaxations lead to the emergence of the $2\sqrt{3}\times 4$, $2\sqrt{3}\times 8$, $4\times 4$, and $4\times 2\sqrt{3}$ structures, all of which represent more stable configurations than their parent phases.

The calculated energy hierarchy reveals that these secondary CDW phases are energetically favored over the first-generation structures, confirming that the initially obtained CDW states correspond to metastable intermediates in the structural transformation pathway. In particular, the $4\times 2\sqrt{3}$ and $2\sqrt{3}\times 4$ phases exhibit nearly identical total energies, indicating that their distinct phonon-driven pathways converge toward a common low-energy region of the CDW energy landscape.

These results demonstrate that the CDW formation in monolayer $VSe_2$ needs to be understood beyond a single phonon instability of the normal phase alone. Instead, the structural evolution is governed by a sequence of competing and interconnected lattice instabilities that drive the system through a cascade of intermediate CDW states toward dynamically stable low-energy configurations. Our work therefore provides a systematic route for identifying hidden metastable phases and ground-state CDW structures in $VSe_2$, and more generally offers a practical strategy for exploring complex CDW phase transitions in low-dimensional materials.

**Acknowledgments**

The research study is supported financially by the National Science and Technology Council (NSTC) of Taiwan (Grant Nos. 110-2112-M-A49-022-MY2 and 113-2112-M-A49-013).

**Author contributions**

Conceptualization, C.C.L., M.C.C., and H.C.H.; Methodology, C.C.L., H.C.H., P.T.C., and Y.T.L.; Investigation, C.C.L., Z.H.L., C.L.L., P.T.C., Y.T.L., and C.C.K.; Visualization, Z.H.L., C.C.L., C.L.L., and P.T.C.; Funding acquisition, C.T.W, M.C.C., and H.C.H.; Project administration, M.C.C. and H.C.H.; Supervision, C.C.L., M.C.C., and H.C.H.; Writing – original draft: P.T.C., Z.H.L., Y.T.L., Y.C.T and C.C.L.; Writing – review and editing, C.T.C., C.L.L., C.C.K., M.K.L., M.C.C., and H.C.H.

## Competing interests

Authors declare that they have no competing interests.